\documentclass[reprint, amsmath, amssymb, amsfonts, aps, a4paper, dvipdfmx]{revtex4-1}
\usepackage[dvipdfmx]{graphicx,color}
\usepackage{bm}
\usepackage{mathtools,amsmath,amsfonts}
\usepackage{ulem}
\usepackage{physics}
\usepackage{comment}

\newcommand{\bB}{\textbf{B}}
\newcommand{\bK}{\textbf{K}}
\newcommand{\bP}{\textbf{P}}
\newcommand{\by}{\bm{y}}

\newcommand{\dfracp}[2]{\dfrac{\partial #1}{\partial #2}}

\begin{document}

\title{Conformation dynamics in asymmetric chain-like three-body bead-spring models}
\author{Yuki Sogo}
\email{ysogo@amp.i.kyoto-u.ac.jp}
%\affiliation{Graduate School of Informatics, Kyoto Uniersity, Kyoto University, Kyoto 606-8501, Japan}

\author{Yoshiyuki Y. Yamaguchi}
\email{yyama@amp.i.kyoto-u.ac.jp}
\affiliation{Graduate School of Informatics, Kyoto University, Kyoto 606-8501, Japan }

\begin{abstract}
  We consider conformation dynamics of a chain-like three-body bead-spring model,
  in which three point masses are connected in series by two springs
  and the conformation is defined by the bending angle between the two springs.
  Previous studies have theoretically shown that
  an unstable (stable) conformation based on the potential function
  can be stabilized (destabilized) by exciting spring vibration
  and stabilization or destabilization depends on amplitudes of vibration modes.
  However, the system was restricted in symmetric cases
  in which the two springs are identical and the masses of the two end beads are identical.
  This symmetry simplifies energy exchange between the vibration modes
  and conformation dynamics accordingly.
  We extend the theory into asymmetric systems.
  This extension can induce nontrivial energy exchange between the modes
  and a corresponding nontrivial conformation dynamics.
\end{abstract}

\maketitle

\section{Introduction}
\label{sec:Introduction}

Stability of conformation in a potential dynamical system is typically discussed 
by analyzing the
given potential energy function, but vibration can change stability.
The Kapitza pendulum \cite{kap1,kap2,Butikov01} is a typical example.
The gravitational potential energy is maximum at the inverted position, but the inverted pendulum
can be stabilized by vibrating its pivot vertically with small amplitude and high frequency.
Another example
is the Paul trap \cite{paul,Leibfried03}.
A quadrupole field creates a saddle point of potential at the center,
but charged particles can be trapped around the saddle by vibrating the quadrupole field.
The above examples provide wide applications in science and enginnering
\cite{Bellman86,Shapiro97,Bukov15,Blatt08,Rosenband08,Kirillov16}.

Stabilization by vibtation is not restricted in nonautonomous systems
in which the vibration is given as an external force
like the above two examples.
Let us consider a planar system consisting of three masses
interacting through the Lenneard-Jones potential \cite{LJ,Fischer23,Lenhard24}.
This system takes the potential minimum at the equilateral triangle conformation.
A linear conformation is unstable since it is at a saddle point of the total potential energy
and is called as a transition state.
The linear conformation can be however stabilized by exciting
vibration of the two distances from the center mass to the two ends \cite{Yamaguchi25}.
Mathematically, the static linear conformation corresponds to an unstable equilibrium point
on the phase space, and it becomes a stable periodic orbit
when the amplitudes of vibration is sufficiently large.

The stabilization mechanism is,
by using chain-like bead-spring models \cite{Rouse53},
theoretically revealed 
in three-body systems \cite{yyy2022} and then extended to $N$-body systems \cite{yyy2023}.
The theory is also applied to explain the stabilization of the transition state
in the above three-body Lennard-Jones system \cite{Yamaguchi25}.
A remarkable nature in autonomous systems
is that the stabilization depends on excited modes of springs.
It is suggested that the mode having the lowest eigenfrequency
contributes to stabilization but the other modes do not.
In a three-body system,
there are two modes so-called the in-phase mode and the antiphase mode,
where the two springs vibrate simultaneously (alternatively)
in the in-phase (antiphase) mode.
The eigenfrequencies depend on conformation
and the eigenfrequency of the in-phase mode is smaller than the antiphase mode
at the straight conformation but larger at the fully bended conformation.
Accordingly,
the straight conformation is stabilized (destabilized)
by the in-phase (antiphase) mode,
while the fully bended conformation is stabilized (destabilized)
by the antiphase (in-phase) mode \cite{yyy2022}.

These theoretical works are restricted in symmetric systems:
All interactions and masses are homogeneous,
or the two springs and the two end masses are identical in three-body chain-like systems.
According to numerical simulations, in a latter system,
excited in-phase and antiphase modes almost keep the mode energy ratio
in spite of conformation change.
This conservation implies that, with the antiphase mode excitation for instance,
the straight conformation is destabilized while the fully bended conformation is stabilized
throughout global conformation dynamics.
Further, an effective potential can be constructed
which is valid in the entire interval of the bending angle \cite{yyy2023}.

The mode energy conservation relies on the degeneracy of the two eigenfrequencies
at a certain conformation.
The degeneracy permits continuous modification of eigenfrequencies
  throughout glocal conformation dynamics
  and the continuity helps to conserve the mode energy ratio.
However, in an asymmetric system, the degeneracy does not occur
in general, and, remembering the change of amplitude relation between the two eigenfrequencies,
  each eigenfrequency has a jump as a function of the bending angle.
The system has two choices: to accept a jump of eigenfrequency or to modify the mode energy ratio.

In an asymmetric system, although the initially excited mode is identified,
stability of the straight or fully bended conformations
may become nontrivial due to nontrivial mode energy transfer as time goes on.
We may reveal a hidden type of conformation dynamics.
It is therefore demanded to extend analyses of the stabilization mechanism
to asymmetric systems and to investigate conformation dynamics.
In this paper, we study chain-like three-body bead-spring systems
characterized by nonidentical masses.
We will answer to the following three questions:
(i) Which vibration mode stabilizes conformation?
(ii) How about the mode energy transfer with emergence of a gap between the two eigenfrequencies?
(iii) Does nondegeneracy of eigenfrequencies bring a new type of dynamics in conformation?

This paper is organized as follows.
We introduce the three-body bead-spring model moving on the plane in Sec.~\ref{sec:model}.
The (de)stabilization is theoretically analyzed in Sec.~\ref{sec:Theory}
by using the multiple-scale analysis \cite{Bender99}
and the averaging method \cite{Krylov34,Krylov47,GuckenHeimer83}.
Conformation dynamics induced by asymmetry is exhibited
through numerical simulations in Sec.~\ref{sec:Numerics}.
Finally, the summary and concluding remarks are provided in Sec.~\ref{sec:Summary}.

\section{model}
\label{sec:model}

\begin{figure}[htbp]
    \centering
    \includegraphics[scale =  0.8]{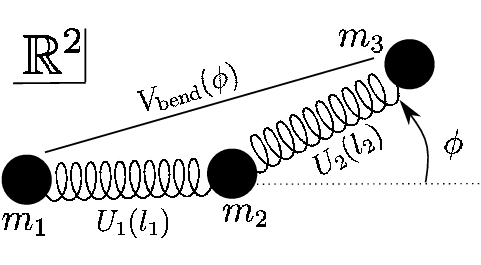}
    \caption{The three-body chain-like bead-spring model moving on a plane.}
    \label{fig:sotu1}
\end{figure}

We consider a three-body bead-spring model moving on a two-dimensional plane.
This model consists of three beads and two springs connecting them.
Let the mass and position vector of the $j$-th bead be
$m_{j}>0$ and $\bm{r}_{j} \in \mathbb{R}^2$, respectively.
The Lagrangian is given by
\begin{equation}
  L = \dfrac{1}{2}\sum_{j=1}^{3} m_{j} \norm{\dot{\bm{r}_{j}}}^{2} - V(\bm{r}_1,\bm{r}_2,\bm{r}_3),
  \label{lag1}
\end{equation}
where $\bm{r}_{j} = (x_{j},y_{j})^{\mathrm{T}},\dot{\bm{r}_{j}} = d\bm{r}_{j}/{dt},\norm{\cdot}$
denotes the Euclidean norm, the superscript $\mathrm{T}$ denotes transposition, and
$V$ represents the total potential energy. We assume that
this system has both translational symmetry and rotational symmetry.
See Fig.~\ref{fig:sotu1} for the model.

We introduce the inner coordinates $\bm{y}=(l_{1},l_{2},\phi,\psi)$,
where $l_{j}=\norm{\bm{r}_{j}-\bm{r}_{j+1}}~(j=1,2)$, $\phi$ is the bending angle,
and $\psi$ is the angle determining the direction of the system.
We included $\psi$ for simplification of computations,
but $\psi$ is a cyclic coordinate from the rotational symmetry and is physically irrelevant.
We will omit it from $\by$ if no confusion occurs.
The Lagrangian in the inner coordinates is written as
\begin{equation}
  L = \dfrac{1}{2} \sum_{\alpha,\beta=1}^{4} B^{\alpha\beta}(\bm{y}) \dot{y}_{\alpha} \dot{y}_{\beta}
  - V(\bm{y}),
  \label{eq:Lagrangian}
\end{equation}
where $B^{\alpha\beta}$ is the $(\alpha,\beta)$ element of symmetric fourth-order square matrix
$\bB$. The explicit forms of $\bB$ and $\bB^{-1}$ are presented in Appendix \ref{sec:matrixB}.
We assume that the total potential energy is expressed as
\begin{equation}
  V = V_{\rm spring}(l_{1},l_{2}) + V_{\rm bend}(\phi),
\end{equation}
where
\begin{equation}
  V_{\rm spring}(l_{1},l_{2}) = U_{1}(l_{1}) + U_{2}(l_{2})
\end{equation}
and that $U_{j}(l_{j})$ takes the minimum value at $l_{j}=l_{j\ast}$.
We note that $U_{j}(l_{j})$ may be a linear or nonlinear spring.

Hereafter, we use the Einstein summation convention for sums:
we take the sum over a suffix if it appears twice in a term.
The Euler-Lagrange equation is expressed as
\begin{equation}
  B^{\alpha \beta}(\bm{y})\ddot{y_{\beta}}
  + \Gamma^{\alpha\beta\gamma}(\bm{y}) \dot{y_{\beta}} \dot{y_{\gamma}}
  = - \dfrac{\partial V}{\partial y_{\alpha}}(\bm{y}),
  \label{eulerlag}
\end{equation}
where
\begin{equation}
  \Gamma^{\alpha\beta\gamma}(\by)
  = \dfracp{B^{\alpha\beta}}{y_{\gamma}}(\by)
  - \dfrac{1}{2} \dfracp{B^{\beta\gamma}}{y_{\alpha}}(\by)
\end{equation}
and $\alpha,\beta,\gamma \in \{1,2,3,4\}$.
The left-hand side in Eq.~\eqref{eulerlag} represents the geodesic flow for the metric $\bB$
and the right-hand side the potential force.

\section{Theory}
\label{sec:Theory}

The main element of stabilization in the system \eqref{eq:Lagrangian}
is emergence of multiscales in time and space.
Fast motion corresponding to vibration of $l_{i}$ provides an effective force
to slow motion of the bending angle $\phi$, as the Kapitza pendulum.
Let the dimensionless parameter $\epsilon$ satisfy $|\epsilon|\ll 1$.
The multiscales are input by introducing the two time scales
\begin{equation}
  t_{0}=t, \qquad t_{1}=\epsilon t,
\end{equation}
and by expanding the inner coordinates as
\begin{equation}
  \begin{split}
    & l_{j}(t_{0},t_{1}) = l_{j\ast} + \epsilon l_{j}^{(1)}(t_{0},t_{1}), \quad (j=1,2) \\
    & \phi(t_{0},t_{1}) = \phi^{(0)}(t_{1}) + \epsilon\phi^{(1)}(t_{0},t_{1}). \\
  \end{split}
\end{equation}
We are interested in slow and large motion of the bending angle
represented by $\phi^{(0)}(t_{1})$. Denoting
\begin{equation}
  \by^{(0)}=(l_{1\ast},l_{2\ast},\phi^{(0)}),
  \quad
  \by^{(1)}=(l_{1}^{(1)},l_{2}^{(1)},\phi^{(1)}),
\end{equation}
we expand the Euler-Lagrange equation \eqref{eulerlag} into a series of $\epsilon$.
Fast small vibration appears in $O(\epsilon)$ and slow large motion in $O(\epsilon^{2})$.
We assume that
\begin{equation}
  V_{\rm bend}(\phi)=\epsilon^{2} V_{\rm bend}^{(2)}(\phi)+O(\epsilon^{3})
\end{equation}
to have a competitive bending potential comparing with the effective force to $\phi^{(0)}(t_{1})$.

\subsection{Fast small vibration in $O(\epsilon)$}

The equations of motion in $O(\epsilon)$ are
\begin{equation}
  \dfracp{{}^{2}\by^{(1)}}{t_{0}^{2}}
  = - \bB(\by^{(0)})^{-1} \bK \by^{(1)},
  \label{eq:Oepsilon0}
\end{equation}
where
\begin{equation}
  \bK = {\rm diag}(k_{1},k_{2},0,0),
  \quad
  k_{j} = U_{j}''(l_{j\ast})>0
\end{equation}
is the Hessian of $V(\by)$ up to $O(\epsilon)$, namely of $V_{\rm spring}(l_{1},l_{2})$.
See Appendix \ref{sec:diagonailzation} for diagonalization of the matrix $\bB(\by^{(0)})^{-1}\bK$.
This matrix has two zero and two nontrivial eigenvalues.
The nontrivial eigenvalues denoted by $\lambda_{1},\lambda_{2}~(\lambda_{1}\leq\lambda_{2})$ are
\begin{equation}
  \lambda_{1}
  = \dfrac{k_{1}M_{3}+k_{2}M_{2}-\sqrt{S}}{2(M_{2}M_{3}-M_{1}^{2})},
  \quad
  \lambda_{2}
  = \dfrac{k_{1}M_{3}+k_{2}M_{2}+\sqrt{S}}{2(M_{2}M_{3}-M_{1}^{2})},
  \label{eq:eigenvalues}
\end{equation}
where
\begin{equation}
  S(\phi^{(0)}) = (k_{1}M_{3}-k_{2}M_{2})^{2}+4k_{1}k_{2}M_{1}^{2}\cos^{2}\phi^{(0)},
  \label{eq:S}
\end{equation}
and
\begin{equation}
  \begin{split}
    & M_{1} = \dfrac{m_{1}m_{3}}{M},
      ~
      M_{2} = \dfrac{m_{1}(m_{2}+m_{3})}{M},
      ~
      M_{3} = \dfrac{m_{3}(m_{2}+m_{1})}{M}, \\
    & M = m_{1} + m_{2} + m_{3}. \\ 
  \end{split}
  \label{eq:M123}
\end{equation}
We refer to the mode corresponding to the eigenvalue $\lambda_{j}$ as mode-$j$.
Note that the in-phase (antiphase) mode is mode-$1$ (mode-$2$) in $|\phi|<\pi/2$,
while it is mode-$2$ (mode-$1)$ in $\pi/2<|\phi|<\pi$.

The degeneracy of the two eigenvalues occurs if $S=0$, which is equivalent with
$k_{1}M_{3}=k_{2}M_{2}$ and $\phi^{(0)}=\pm\pi/2$.
Focusing on the case $k_{1}=k_{2}$,
the degeneracy condition is read as $m_{1}=m_{3}$, which implies symmetry of the masses.
A slight discrepancy between $m_{1}$ and $m_{3}$ breaks the degeneracy,
and the gap between the eigenvalues increases as asymmetry is enhanced.

\subsection{Slow large motion in $O(\epsilon^{2})$}

The equations of motion for $\by^{(0)}(t_{1})$ is given in $O(\epsilon^{2})$,
which are read as
\begin{equation}
  \begin{split}
    & B^{\alpha\beta}(\by^{(0)})
      \left( \dfrac{d^{2}y_{\beta}^{(0)}}{dt_{1}^{2}}
      + 2 \dfracp{{}^{2}y_{\beta}^{(1)}}{t_{0}\partial t_{1}} \right)
      + \Gamma^{\alpha\beta\gamma}(\by^{(0)}) (\dot{y}_{\beta})^{(1)} (\dot{y}_{\gamma})^{(1)}  \\
    & + \dfracp{B^{\alpha\beta}}{y_{\gamma}}(\by^{(0)})
      \dfracp{{}^{2}y_{\beta}^{(1)}}{t_{0}^{2}} y_{\gamma}^{(1)}
      = - \dfracp{V_{\rm bend}^{(2)}}{y_{\alpha}}(\phi^{(0)}),
      \label{eq:Oepsilon2}
  \end{split}
\end{equation}
where
\begin{equation}
  (\dot{y}_{\beta})^{(1)}
  = \dfrac{dy_{\beta}^{(0)}}{dt_{1}}(t_{1}) + \dfracp{y_{\beta}^{(1)}}{t_{0}}(t_{0},t_{1}).
\end{equation}
The left-hand side in Eq.~\eqref{eq:Oepsilon2} contains fast vibration of $\by^{(1)}(t_{0},t_{1})$
and we perform the averaging over the fast timescale $t_{0}$.
The linear terms with respect to $\by^{(1)}$ vanish by the averaging,
but the quadratic terms survive.
The averaged equations are
\begin{equation}
  \begin{split}
    & B^{\alpha\beta}(\by^{(0)}) \dfrac{d^{2}y_{\beta}^{(0)}}{dt_{1}^{2}}
    + \Gamma^{\alpha\beta\gamma}(\by^{(0)}) \dfrac{dy_{\beta}^{(0)}}{dt_{1}} \dfrac{dy_{\gamma}^{(0)}}{dt_{1}} \\
    & = - \dfracp{V_{\rm bend}^{(2)}}{y_{\alpha}} + F_{\rm eff}^{\alpha},
  \end{split}
  \label{eq:Oepsilon2-ave}
\end{equation}
where
\begin{equation}
  F_{\rm eff}^{\alpha}
  = \dfrac{1}{2} {\rm Tr}\left(
    \dfracp{\bB}{y_{\alpha}}(\by^{(0)})
    \left\langle \dfracp{\by^{(1)}}{t_{0}} \left( \dfracp{\by^{(1)}}{t_{0}} \right)^{\rm T}
    \right\rangle
  \right)
  \label{eq:Feff}
\end{equation}
and $\left\langle\cdots\right\rangle$ denotes the average over $t_{0}$.
Comparing Eq.~\eqref{eq:Oepsilon2-ave} with Eq.~\eqref{eulerlag},
we find an extra term $F_{\rm eff}^{\alpha}$,
which is the effective force induced by averaging.
An effective stationary state and its stability are obtained by analyzing the right-hand side.
The slow and large motion of the bending angle, $\phi^{(0)}(t_{1})$,
is described by setting $\alpha=3$.
We note that $F_{\rm eff}^{3}=0$ at $\phi^{(0)}=0$ and $\pi$.

The effective force $F_{\rm eff}^{\alpha}$ depends on $\phi^{(0)}$ and normal mode energy.
Indeed, the matrix $\bB(\by^{(0)})$ and the eigenvalues of normal modes depend on $\phi^{(0)}$.
Moreover, the normal modes in $\by^{(1)}$ are determined by their amplitudes and initial phases,
and the initial phases are averaged out but the amplitudes are not.
This observation explains why stabilization depends on conformation $\phi$
and energy of excited modes, i.e. the amplitudes.

\section{Numerics}
\label{sec:Numerics}

In this section we perform numerical simulations,
whose setting is introduced in Sec.~\ref{sec:setting}.
Stability of the straight conformation $\phi=0$ is examined in Sec.~\ref{sec:Stabilization}
without the bending potential
to verify the theory developed in Sec.~\ref{sec:Theory}.
Then, we observe conformation dynamics induced by asymmetry
in Secs.~\ref{sec:conformation-dynamics-without} and \ref{sec:conformation-dynamics-with}
without and with the bending potential respectively.

\subsection{Setting}
\label{sec:setting}

We perform molecular dynamics simulations using the 4th-order symplectic integrator \cite{si4} with the time step $\Delta t = 0.005$.
To use an explicit algorithm, we come back to the Cartesian coordinates.
The Hamiltonian derived from the Lagrangian \eqref{lag1} is
\begin{equation}
  H
  = \dfrac{1}{2}\sum_{j =1}^{3} \dfrac{\norm{\bm{p}_{j}}^{2}}{m_{j}}
  + V(\bm{r}_{1},\bm{r}_{2},\bm{r}_{3}),
\end{equation}
where $\bm{p}_{j}$ is the conjugate momentum vector to the position vector $\bm{r}_{j}$.
The parameters $m_1 = 1,k_1 = k_2 = 10$ and $l_{1*}=l_{2*}=1$ are fixed throughout simulations.
The spring potential $V_{\text{spring}}$ is
\begin{equation}
  V_{\text{spring}} = \sum_{j=1}^2 \dfrac{k_j}{2}(\norm{\bm{r}_{j+1}-\bm{r}_{j}}-l_{j*})^2.
\end{equation}
Linearity of springs is not essential at least theoretically,
since $V_{\rm spring}$ appears only as the Hessian $\bK$ in $O(\epsilon)$.
We set the bending potential $V_{\text{bend}}$
\begin{equation}
  V_{\text{bend}} = h\cos \phi,
  \qquad
  h = O(\epsilon^{2}).
  \label{eq:Vbend}
\end{equation}
Note that $\phi=\pm\pi$ is not a singular point of $V_{\rm bend}$,
and the fully bended conformation is allowed.
In other words, collision between the two end masses does not affect dynamics.

The initial conditions are given as follows.
All the particles are put as they form the straight conformation,
and all the momenta are zero, $\bm{p}_{j}=\bm{0}~(j=1,2,3)$.
Then, we apply small displacements to $\bm{r}_{1}$ and $\bm{r}_{3}$
for exciting the normal modes,
and to $\bm{r}_{2}$ for modifying the straight conformation.
Denoting the displacements to $\bm{r}_{j}$ by $\delta\bm{r}_{j}$, they are
\begin{equation}
  \delta\bm{r}_{1} = - \epsilon
  \begin{pmatrix}
    \delta l_{1} \\ 0
  \end{pmatrix},
  \quad
  \delta\bm{r}_{2} =
  \begin{pmatrix}
    0 \\ 0.01
  \end{pmatrix},
  \quad
  \delta\bm{r}_{1} = \epsilon
  \begin{pmatrix}
    \delta l_{2} \\ 0
  \end{pmatrix}.
\end{equation}
Excited modes and their strength are controlled by $\delta l_{1}$ and $\delta l_{2}$:
$(\delta l_{1})(\delta l_{2})>0$ for the in-phase mode
and $(\delta l_{1})(\delta l_{2})<0$ for the antiphase mode.
We denote strength of the excited modes by
\begin{equation}
  A = \epsilon\sqrt{(\delta l_{1})^{2}+(\delta l_{2})^{2}}.
  \label{eq:A}
\end{equation}
In the latter computations the value of $\epsilon$ is included in the value of $A$.

\subsection{Stabilization and its critical point without bending potential}
\label{sec:Stabilization}

It has been reported in the symmetric system without bending potential ($h=0$)
\cite{yyy2022}
that the straight conformation $\phi=0$ is stabilized by exciting mode-$1$
and is destabilized by mode-$2$.
However, stabilization holds if the normal mode energy ratio $a$
is smaller than the critical point $a_{\rm c}$, where
\begin{equation}
  a = \dfrac{E_{2}}{E_{1}+E_{2}}
\end{equation}
and $E_{j}$ is the $j$-th normal mode energy.
The critical point is $a_{\rm c}=3/4$ in the identical mass case for instance.
We compute the critical point with varying the masses $m_{2}$ and $m_{3}$ for the fixed $m_{1}=1$.

\begin{figure}[ht]
    \includegraphics[width = 8cm]{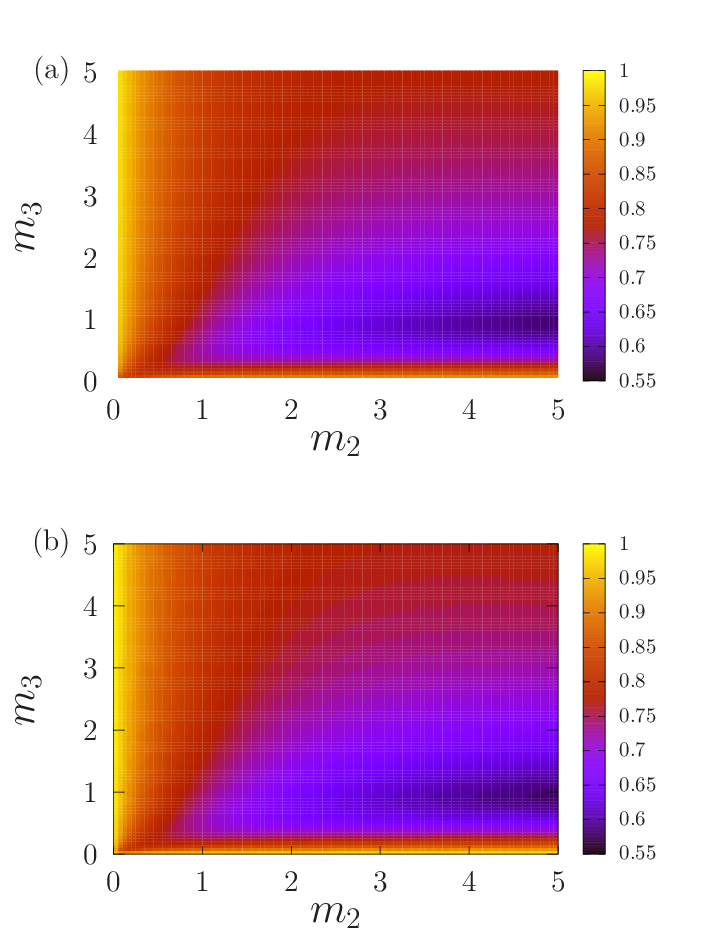}
    \caption{Heat map of critical point $a_{\rm c}$ of stabilization on the plane $(m_{2},m_{3})$.
    $m_{1}=1$.
    (a) Theory. (b) Numerical simulations.
    The color represents the value of $a_{\rm c}$,
      up to which the straight conformation is stabilized.}
      A = 0.2.
  
    \label{fig:critical}
\end{figure}

The theoretical prediction is compared with numerically obtained critical point
in Fig.~\ref{fig:critical}.
The theoretical diagram is obtained by analyzing the effective force \eqref{eq:Feff}.
Numerically, stabilization is judged if $|\phi(t)|<\phi_{\rm th}$ holds
in the time interval $t\in [0,2000]$ with $\phi_{\rm th}=0.2\pi$.
We have two conclusions from Fig.~\ref{fig:critical}.
First, the theory is in good agreement with numerical simulations.
Second, stabilization is enhanced if $m_{3}$ becomes far from $1$, namely more asymmetric,
or $m_{2}$ becomes small.

\subsection{Conformation dynamics without bending potential}
\label{sec:conformation-dynamics-without}

%\begin{widetext}
\begin{figure*}[htp]
    \includegraphics[width = 16cm]{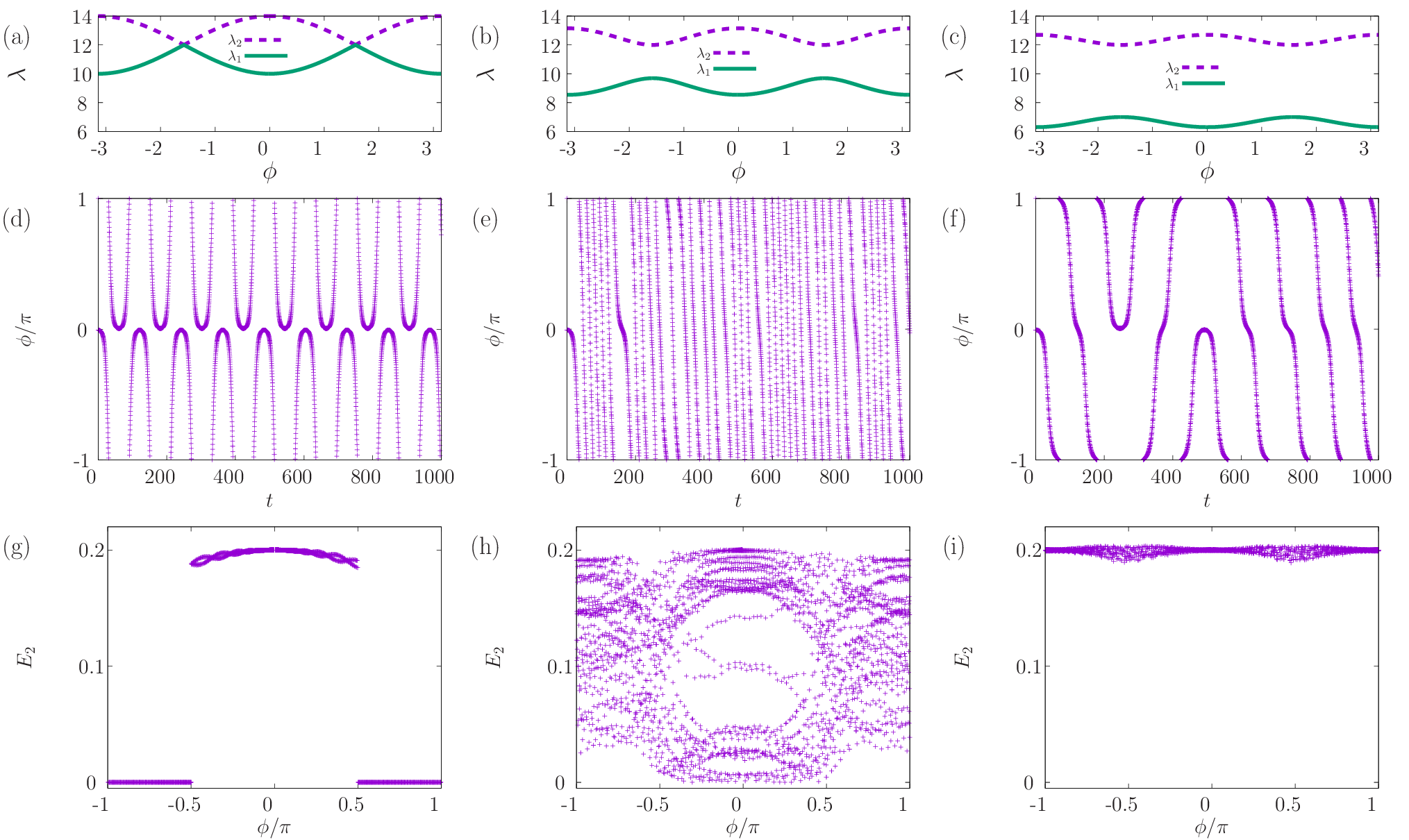}
    \caption{
      Three types of conformation dynamics.
      (a,b,c) Nontrivial eigenvalues of $\bB^{-1}\bK$ as functions of the bending angle $\phi$.
      (d,e,f) Temporal evolution of the bending angle $\phi$.
      (g,h,i) Mode-2 energy $E_{2}$ as a function of $\phi/\pi$.
      $m_{1}=1$ and $m_{2}=5$ are fixed,
      and $m_{3}=1$ (a,d,g), $1.3$ (b,e,h), and $2$ (c,f,i). $A=0.2$.
    }
    \label{fig3new}
\end{figure*}
%\end{widetext}

We reveal qualitative difference caused by asymmetry.
With fixing $m_{1}=1$, $m_{2}=5$, we modify $m_{3}$ as $m_{3}=1, 1.3,$ and $2$.
We will classify conformation dynamics into three types depending on asymmetry.
The mode energy ratio $a$ is set as $a=1$ to observe conformation dynamics
by inducing instability to the straight conformation ($\phi=0$).

First, we observe $\phi$ dependence of the two eigenvalues $\lambda_{j}~(j=1,2)$.
The gap between the two eigenvalues reaches zero in the symmetric case
[$m_{3}=1$ in Fig.~\ref{fig3new}(a)]
and is enhanced as asymmetry increases
[$m_{3}=1.3$ in Fig.~\ref{fig3new}(b) and $m_{3}=3$ in Fig.~\ref{fig3new}(c)].

Second, temporal evolution of the bending angle $\phi$
is shown in Figs.~\ref{fig3new}(d), (e), and (f).
We focus on the relation between stability and excited modes at
two stationary states $\phi=0$ and $\pi$.
In the symmetric case, Fig~\ref{fig3new}(d), the straight conformation $\phi=0$ is repulsed
while $\phi(t)$ is rather accelerated at $\phi=\pi$.
We therefore conclude that $\phi=0$ is unstable and $\phi=\pi$ is stable.
This observation is consistent with the mode-$2$ energy which contributes to destabilization:
$E_{2}$ is large in $|\phi|<\pi/2$ but is almost zero in $\pi/2<|\phi|<\pi$
[Fig.~\ref{fig3new}(j)].
Note that this $\phi$ dependence of $E_{2}$ corresponds to conservation of
the antiphase mode energy, which becomes mode-$1$ in $\pi/2<|\phi|<\pi$.
The above nature is broken by asymmetry.
With $m_{3}=1.3$, the straight conformation $\phi=0$ may be stable or unstable
depending on the visiting timing [Fig.~\ref{fig3new}(e)].
Accordingly, $E_{2}$ may be large or small in the whole range of $\phi$ [Fig.~\ref{fig3new}(h)].
Further increasing $m_{3}$, both $\phi=0$ and $\pi$ are unstable [Fig.~\ref{fig3new}(f)]
with $m_{3}=2$, and $E_{2}$ is always large [Fig.~\ref{fig3new}(i)].
Temporal evolution of $E_{2}$ reflects facility of mode energy exchange,
  which gets harder as the gap between $\lambda_{1}$ and $\lambda_{2}$ becomes larger.

Change of stability at $\phi=0$ or $\phi=\pi$ can be explained as follows.
The mode energy is concentrated in mode-$2$ at the initial time in the straight conformation $\phi=0$.
In a symmetric case this mode energy is almost completely transferred to mode-$1$
at $\phi=\pm\pi/2$ due to the degeneracy of the two eigenvalues
and $\phi=\pi$ becomes stable by excitation of mode-$1$.
However, the transfer is not perfect for $m_{3}=1.3$ and is impossible for $m_{3}=2$
due to emergence of gaps between the two eigenvalues.
In the latter case in particular, $\phi=\pi$ becomes unstable
while it is stable in a symmetric case ($m_{3}=1$).

As shown in Fig.~\ref{fig3new},
conformation dynamics $\phi(t)$ depends on the gap of the eigenvalues induced by $m_{3}$.
It also depends on strength of mode excitation $A$.
To classify the $(m_{3},A)$ plane into the three types of dynamics shown in Fig.~\ref{fig3new},
we introduce two quantities.
  Dividing the whole interval of $\phi\in(-\pi,\pi]$ into $200$ bins,
  we define the two quantities by
\begin{equation}
  \begin{split}
    & M_{\rm var} = \mathbb{E}[ \mathbb{V}_{i}[E_{1}] ],
      \qquad
      V_{\rm ave} = \mathbb{V}[ \mathbb{E}_{i}[E_{1}] ], \\
  \end{split}
  \label{eq:Mvar-Vave}
\end{equation}
where $\mathbb{E}_{i}$ and $\mathbb{V}_{i}$ represent the average and variance
over time under the condition that $\phi(t)$ is in the $i$-th bin,
and $\mathbb{E}$ and $\mathbb{V}$ the average and variance over the $200$ bins.
$M_{\rm var}$ is small when $E_{1}$ is described as a graph of $\phi$
as Figs.~\ref{fig3new}(g) and (i),
and moreover $V_{\rm ave}$ is small when the graph is almost constant
as Fig.~\ref{fig3new}(i), although the panels are for $E_{2}$.

%\begin{widetext}
\begin{table*}
  \begin{center}
    \caption{Classification criterion on the $(m_{3},A)$ plane into three zones.
      The initial conformation is the straight conformation $\phi=0$.
      The initially excited mode is mode-$2$.
      This setting induces destabilization of $\phi=0$
      and stabilization of $\phi=\pi$ when the system is symmetric.
      See the text for the definitions of $R_{M},R_{V}$, and $R_{\rm th}$.
      Energy exchange represents exchange of mode energy between mode-$1$ and mode-$2$.
      Stability of $\phi=\pi$ is from observation. $k_{1}=k_{2}=10$.
    }
    \begin{tabular}{c|c|c|c|c|c}
      Zone & Masses & Definition & Eigenvalues & Energy exchange & Stability of $\phi = \pi$\\ \hline
      I & Symmetric & $R_{M}<R_{\rm th}$ and $R_{V}>R_{\rm th}$ & Fig.~\ref{fig3new}(a) & Large & Stable \\ %$V_{\bf eff}$ is local minimum\\
      II & \bf{Intermediate} & $R_{M}>R_{\rm th} $ & Fig.~\ref{fig3new}(b) & Depend on timing & Depending on timing \\ %$l_1(t),l_2(t)$&Depend on $l_1(t),l_2(t)$\\
      III & \bf{Asymmetric} & $R_{M}<R_{\rm th}$ and $R_{V}<R_{\rm th}$ & Fig.~\ref{fig3new}(c)& None & Unstable \\ %$V_{\bf eff}$ is local maximum
    \end{tabular}
  \end{center}
  \label{tab:table11}
\end{table*}
%\end{widetext}

\begin{figure}[htbp]
    \includegraphics[width = 8cm]{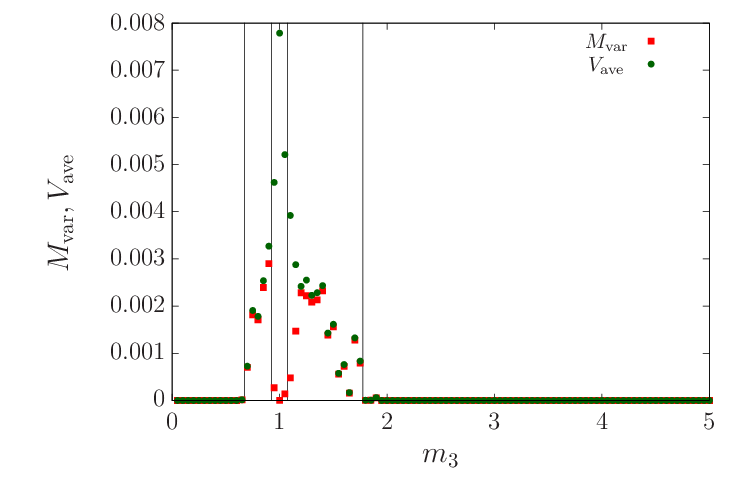}
    \caption{Average of variances $M_{\rm var}$ (red squares)
        and variance of averages $V_{\rm ave}$ (green circles)
      of the mode-$1$ energy as a function of $m_{3}$.
      See Eq.~\eqref{eq:Mvar-Vave} and the text for their definitions.
      $A=0.2$. $m_{1}=1, m_{2}=5$. $k_{1}=k_{2}=10$. $t \in [0,1000]$.}
    \label{dispersion}
\end{figure}

An example of $M_{\rm var}$ and $V_{\rm ave}$ is reported in Fig.~\ref{dispersion}
for a fixed $A=0.2$ as a function of $m_{3}$.
As expected, $M_{\rm var}$ is small around the symmetric mass $m_{3}=1$
  and far from $m_{3}=1$, and $V_{\rm ave}$ is small far from $m_{3}=1$.

The classification of the $(m_{3},A)$ plane is carried out as follows.
Let $M_{\rm var}^{\rm max}$ and $V_{\rm ave}^{\rm max}$ be the maximum values over $m_{3}$ of
$M_{\rm var}$ and $V_{\rm ave}$ respectively for a given $A$.
From Fig.~\ref{dispersion}, $V_{\rm ave}^{\rm max}\simeq 0.00778$ for $A=0.2$ for example.
We introduce the ratios $R_{M}=M_{\rm var}/M_{\rm var}^{\rm max}$
and $R_{V}=V_{\rm ave}/V_{\rm ave}^{\rm max}$
as well as the threshold $R_{\rm th}=0.1$.
A point $(m_{3},A)$ is classified into Zone I when $R_{M}<R_{\rm th}$ and $R_{V}>R_{\rm th}$,
Zone II when $R_{M}>R_{\rm th}$,
and Zone III when $R<R_{M}$ and $R_{V}<R_{\rm th}$.
See Table I for a summary of the three zones.

\begin{figure}[htbp]
    \includegraphics[width = 8cm]{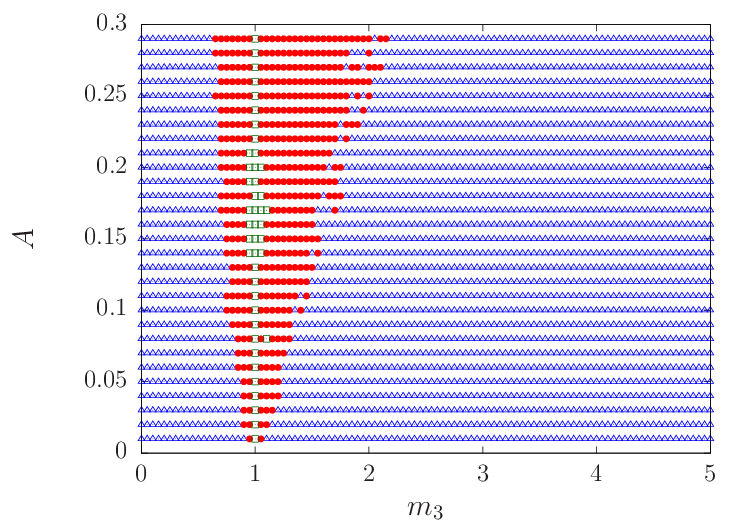}
    \caption{Phase diagram on the plane $(m_{3},A)$ with $m_{1}=1, m_{2}=5$ and $k_{1}=k_{2}=10$.
      Zone I (green squares), Zone II (red circles), and Zone III (blue triangles).
    }
    \label{phasediagram}
\end{figure}

The phase diagram is shown in Fig.~\ref{phasediagram}.
The green area, Zone I, is the symmetric case reported in a previous work \cite{yyy2022}.
Exploring an asymmetric area, $m_{3}\neq 1$,
we find two new regions of Zone II and Zone III,
in which conformation dynamics is qualitatively new.
This phase diagram is our first main contribution in this paper.

\subsection{Conformation dynamics with bending potential}
\label{sec:conformation-dynamics-with}

We demonstrate novelty of the two new zones, Zones II and III, by introducing the bending potential
$V_{\rm bend}(\phi)$ defined in Eq.~\eqref{eq:Vbend} with $h=-0.001$.
This bending potential makes the straight conformation $\phi=0$ stable
and the fully bended conformation $\phi=\pi$ unstable.
We observe conformation dynamics by starting from the straight conformation
with excitation of mode-$2$,
which destabilizes $\phi=0$ and stabilizes $\phi=\pi$ in the symmetric case.
A highlight of this section is stability at $\phi=\pi$,
which is modified by asymmetry.

\begin{figure*}[htbp]
    \includegraphics[width = 16cm]{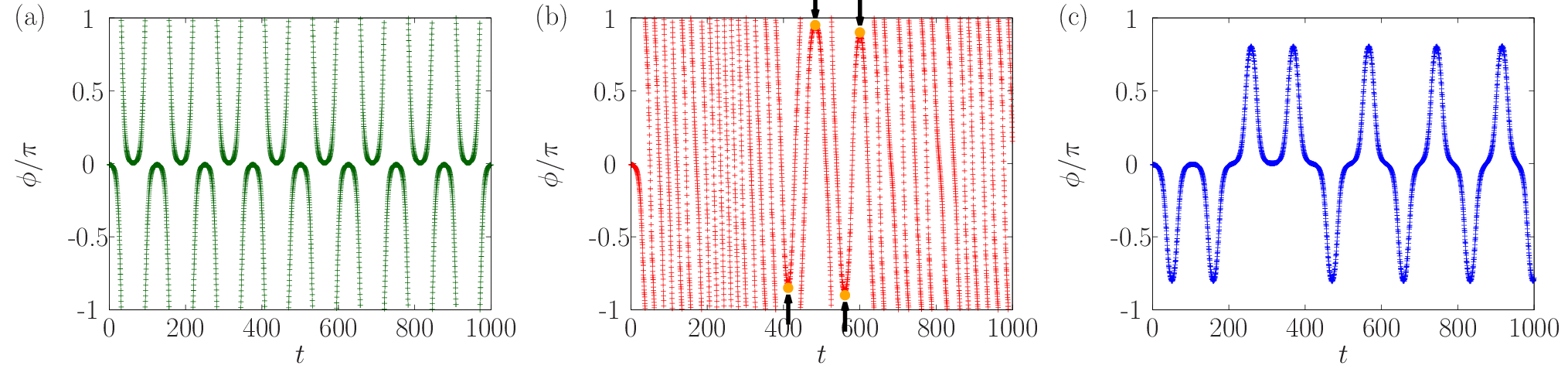}
    \caption{Temporal evolution of the bending angle $\phi$.
      The bending potential $V_{\text{bend}} = -0.001 \cos\phi$.
      (a) $m_{3}=1$ in Zone I.
      (b) $m_{3}=1.3$ in Zone II. Black arrows mark turning points.
      (c) $m_{3}=2$ in Zone III.
      $m_{1} = 1, m_{2} = 5$, $k_{1}=k_{2}=10$. $A=0.2$.
          }
    \label{fig6new}
\end{figure*}

Temporal evolution of the bending angle $\phi$ is shown in Fig.~\ref{fig6new}.
As the previous work, in the symmetric case with $m_{3}=1$,
the straight conformation $\phi=0$ is unstable
and the fully bended conformation $\phi=\pi$ is stable,
where stability is judged by observing $\phi(t)$ as done in Fig.~\ref{fig3new}.

In an asymmetric case with $m_{3}=1.3$,
stability at $\phi=\pi$ is sometimes modified from stable to unstable
at the points marked by black arrows in Fig.~\ref{fig6new}(b),
since the orbit is repulsed from $\phi=\pi$.
The instability depends on the visiting timing at $\phi=\pi$ however.
Further increasing $m_{3}$ to $m_{3}=2$,
$\phi=\pi$ becomes always unstable as shown in Fig.~\ref{fig6new}(c).
This modification of stability is understood from the mode energy transfer
reported in Figs.~\ref{fig3new}(g,h,i):
Complete transfer from mode-$2$ to mode-$1$ makes $\phi=\pi$ stable for the symmetric case
of $m_{3}=1$,
while zero transfer makes it unstable for an asymmetric case of $m_{3}=2$.
This application is our second main contribution
to conformation dynamics caused by asymmetry.

%\begin{widetext}
%\end{widetext}

\begin{figure}[htbp]
  \includegraphics[width=8cm]{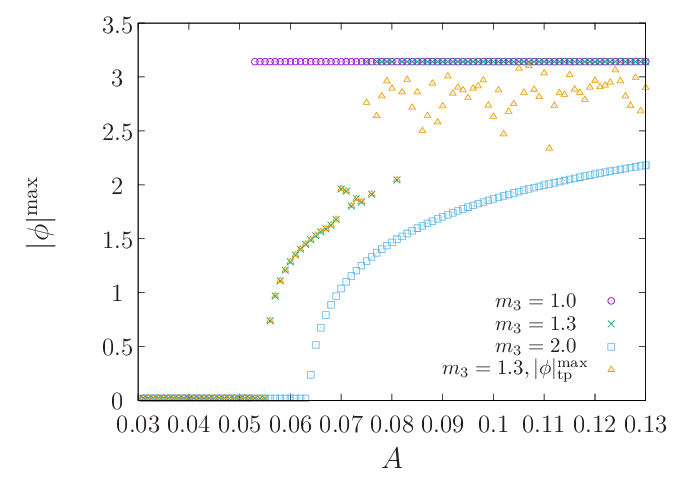}
  \caption{The maximum value $|\phi|^{\rm max}$
    and the maximum turning point $|\phi|_{\rm tp}^{\rm max}$
    of the bending angle $\phi$ in $t\in [0,1000]$
    as functions of initial modification $A$ of spring lengths.
    The maximum value $|\phi|^{\rm max}$ is computed
    for $m_{3}=1$ (purple circles),
    $m_{3}=1.3$ (green crosses),
    and $m_{3}=2$ (blue squares).
    The maximum turning point $|\phi|_{\rm tp}^{\rm max}$
    is computed for $m_{3}=1.3$ in Zones II and III (orange triangles).
    $m_{1}=1$, $m_{2}=5$, $k_{1}=k_{2}=10$.
    }
    \label{fig7new}
\end{figure}

Qualitative differences among the three zones are further clarified
by observing the maximum value and the maximum turning point of $\phi$
under the same setting with Fig.~\eqref{fig6new}.
The maximum value of $|\phi|$ is computed in $t\in [0,1000]$ and denoted by $|\phi|^{\rm max}$,
where $\phi$ is defined in the interval $(-\pi,\pi]$.
A turning point is defined as the point where $\dot{\phi}=0$ in Zones II and III.
Among them, the maximum value of $|\phi|$ is denoted by $|\phi|_{\rm tp}^{\rm max}$.
The two maximum values are picked up in $t\in [0,1000]$
and reported in Fig.~\ref{fig7new} as functions of the initial modification $A$
of spring lengths.
For $m_{3}=1$ (Zone I) the straight conformation $\phi=0$ is stable up to a certain value of $A$,
since the potential force by $V_{\rm bend}$ dominates the effective force $F_{\rm eff}^{3}$.
Once the conformation becomes unstable due to a strong enough effective force,
$\phi$ moves on the whole interval
as shown in Fig.~\ref{fig6new}(a) since the fully bended conformation $\phi=\pi$ is stabilized,
and $|\phi|^{\rm max}$ has a jump accordingly.
For $m_{3}=2.0$ (Zone III), the fully bended conformation $\phi=\pi$ is destabilized
and $|\phi|$ does not reach to the maximum value $\pi$.
For $m_{3}=1.3$, a continuous growing up of $|\phi|^{\rm max}$ is observed
in a small $A$ regime, since it is in Zone III.
In a large $A$ regime, the system enters into Zone II (see Fig.~\ref{phasediagram}),
stability of $\phi=\pi$ changes by time,
and $|\phi|^{\rm max}$ can reach the maximum value $\pi$.
Nevertheless, turning points exist and $|\phi|_{\rm tp}^{\rm max}$ cannot reach $\pi$.

\section{Summary}
\label{sec:Summary}

We have extended conformation stabilization theory by vibration in asymmetric systems.
As suggested in previous works \cite{yyy2022,yyy2023,Yamaguchi25},
the lowest frequency normal mode stabilizes conformation,
while the other one destabilizes.
However, mode energy transfer is modified from complete transfer
to a partial transfer or the complete nontransfer.
This change of energy transfer brings two new types of conformation dynamics,
which appear in so-called Zone II and Zone III on a phase diagram.

Our findings open the door not only to develop theories but also
to design a desired conformation dynamics.
For instance, the investigated three-body system shows scissor-like movements
  in Zone III, which could not be realized in symmetric systems.
The design demands to solve an inverse problem, and it is left as a future work.
Further, effects of dimensionality, noise, and friction are interesting topics to study.

\acknowledgements
Y.Y.Y. acknowledges the support of JSPS KAKENHI Grant No. 21K03402.

\appendix

\section{The matrix $\bB$ and its inverse matrix}
\label{sec:matrixB}

We divide the $4\times 4$ symmetric matrix $\bB$ into four $2\times 2$ matrices as
\begin{equation}
    \bB = \begin{pmatrix}
    \bB_{ll} & \bB_{l\phi}  \\
    \bB_{\phi l} & \bB_{\phi \phi}
  \end{pmatrix}.
\end{equation}
The explicit form of each block is
\begin{equation}
  \bB_{ll} =
  \begin{pmatrix}
    M_2 & M_1 \cos \phi  \\
    M_1 \cos \phi & M_3 \\
  \end{pmatrix},
  \label{eq:Bll}
\end{equation}
\begin{equation}
  \bB_{l\phi} = \dfrac{1}{2} M_{1} \sin\phi
  \begin{pmatrix}
    -l_{2} & -l_{2}  \\
    -l_{1}  & l_{1} \\
  \end{pmatrix},
\end{equation}
$\bB_{\phi l}=\bB_{l\phi}^{\rm T}$, and
% \begin{equation}
%   \bB_{\phi l} =
%   \begin{pmatrix}
%     -\dfrac{1}{2}M_1l_2\sin\phi & -\dfrac{1}{2}M_1l_1\sin\phi  \\
%     -\dfrac{1}{2}M_1l_2\sin\phi & \dfrac{1}{2}M_1l_1\sin\phi
%   \end{pmatrix},
% \end{equation}
\begin{equation}
  \bB_{\phi\phi} =
  \begin{pmatrix}
    B_{33} & -\dfrac{1}{4}(M_2l_1^2-M_3l_2^2)  \\
    -\dfrac{1}{4}(M_2l_1^2-M_3l_2^2)  & B_{44}
  \end{pmatrix},
\end{equation}
where
\begin{equation}
  \begin{split}
    & B_{33}=\dfrac{1}{4}(M_2l_1^2+M_3l_2^2)-\dfrac{1}{2}M_1l_1l_2\cos\phi, \\
    & B_{44}=\dfrac{1}{4}(M_2l_1^2+M_3l_2^2)+\dfrac{1}{2}M_1l_1l_2\cos\phi. \\
  \end{split}
\end{equation}
$M_{j}~(j=1,2,3)$ and $M$ are defined in Eq.~\eqref{eq:M123}.

The inverse matrix $\bB^{-1}$ is
\begin{equation}
  \bB^{-1} = \dfrac{1}{M_2M_3-M_1^2}
  \begin{pmatrix}
    \tilde{\bB}_{ll} & \tilde{\bB}_{l\phi}  \\
    \tilde{\bB}_{\phi l} & \tilde{\bB}_{\phi \phi}
  \end{pmatrix}
\end{equation}
where
\begin{equation}
  \tilde{\bB}_{ll} =
  \begin{pmatrix}
    M_3 & M_1 \cos \phi  \\
    M_1 \cos \phi & M_2
  \end{pmatrix},
\end{equation}
\begin{equation}
  \tilde{\bB}_{l\phi} = M_{1} \sin\phi
  \begin{pmatrix}
    1/l_{2} & 1/l_{2} \\
    1/l_{1} & -1/l_{1} \\
  \end{pmatrix},
\end{equation}
$\tilde{\bB}_{\phi l}=\tilde{\bB}_{l\phi}^{\rm T}$, and
\begin{equation}
    \tilde{\bB}_{\phi \phi} = \begin{pmatrix}
        \tilde{B}_{33} &\dfrac{M_2l_1^2-M_3l_2^2}{l_1^2l_2^2}  \\
    \dfrac{M_2l_1^2-M_3l_2^2}{l_1^2l_2^2}  & \tilde{B}_{44}
    \end{pmatrix},
\end{equation}
with
\begin{equation}
  \begin{split}
    & \tilde{B}_{33}=\dfrac{M_2l_1^2+M_3l_2^2+2M_1l_1l_2 \cos \phi }{l_1^2l_2^2}, \\
    & \tilde{B}_{44}=\dfrac{M_2l_1^2+M_3l_2^2-2M_1l_1l_2 \cos \phi }{l_1^2l_2^2}. \\
  \end{split}
\end{equation}

\section{Diagonalization of the matrix $\bB^{-1}\bK$}
\label{sec:diagonailzation}

The matrix $\bB^{-1}(\phi)\bK$ is diagonalized by a regular matrix $\bP$
consisting of four eigenvectors
\begin{equation}
  \bP(\phi) = (\bm{p}_{\lambda_1},\bm{p}_{\lambda_2},\bm{p}_{\phi},\bm{p}_{\psi}),
  \label{matP}
\end{equation}
where $\bm{p}_{\lambda_{j}}$ is the eigenvector corresponding to the eigenvalue $\lambda_{j}$
and the other two vectors to the zero eigenvalues.
Explicit forms of the eigenvectors are
\begin{equation}
  \bm{p}_{\lambda_{1}} =
  \begin{pmatrix}
    2 k_{2}M_{1}\cos\phi \\
    -(k_{1}M_{3}-k_{2}M_{2})+\sqrt{S(\phi)} \\
    D_{1} \\
    D_{2} \\
  \end{pmatrix},
\end{equation}
\begin{equation}
  \bm{p}_{\lambda_{2}} =
  \begin{pmatrix}
    2 k_{2}M_{1}\cos\phi \\
    -(k_{1}M_{3}-k_{2}M_{2})-\sqrt{S(\phi)} \\
    D_{3} \\
    D_{4} \\
  \end{pmatrix},
\end{equation}
\begin{equation}
  \bm{p}_{\phi} =
  \begin{pmatrix}
    0 \\
    0 \\
    \dfrac{1}{4}(M_{2}l_{1}^{2}+M_{3}L_{2}^{2}) + \dfrac{1}{2}M_{1}l_{1}l_{2}\cos\phi \\
    \dfrac{1}{4}(M_{2}l_{1}^{2}-M_{3}l_{2}^{2}) \\
  \end{pmatrix},
\end{equation}
and
\begin{equation}
  \bm{p}_{\psi} =
  \begin{pmatrix}
    0 \\
    0 \\
    0 \\
    1 \\
  \end{pmatrix}.
\end{equation}
%\red{\it (Are they normalized?)}
Here, $S(\phi)$ is defined in Eq.~\eqref{eq:S}.
$D_{j}~(j=1,2,3,4)$ are defined by
\begin{equation}
  \begin{split}
    & D_{1} = 2k_{2}M_{1}\sin\phi \dfrac{d_{1}(\phi)+d_{2}[-T+\sqrt{S(\phi)}]}{d_{3}(\phi)}, \\
    & D_{2} = 2k_{2}M_{1}\sin\phi \dfrac{d_{1}(\phi)-d_{2}[-T+\sqrt{S(\phi)}]}{d_{3}(\phi)}, \\
    & D_{3} = 2k_{2}M_{1}\sin\phi \dfrac{d_{1}(\phi)-d_{2}[T+\sqrt{S(\phi)}]}{d_{4}(\phi)}, \\
    & D_{4} = 2k_{2}M_{1}\sin\phi \dfrac{d_{1}(\phi)+d_{2}[T+\sqrt{S(\phi)}]}{d_{4}(\phi)},
  \end{split}
\end{equation}
and
\begin{equation}
  \begin{split}
    & T = k_{1}M_{3}-k_{2}M_{2}, \\
    & d_{1}(\phi) = \dfrac{2k_{1}M_{1}\cos\phi}{l_{2}}, \\
    & d_{2} = \dfrac{1}{l_{1}}, \\
    & d_{3}(\phi) = k_{1}M_{3}+k_{2}M_{2} + \sqrt{S(\phi)}, \\
    & d_{4}(\phi) = k_{1}M_{3}+k_{2}M_{2} - \sqrt{S(\phi)}. \\
  \end{split}
\end{equation}

\end{document}